\documentclass[times, twoside]{zHenriquesLab-StyleBioRxiv}
\usepackage{blindtext}
\usepackage{multirow}
\usepackage{makecell}
\usepackage{lipsum}
\usepackage{array}
\usepackage[table]{xcolor}
\usepackage{makecell}
\leadauthor{Bourgade}

\begin{document}

\title{Robust Pan-Cancer Mitotic Figure Detection with YOLOv12}
 \shorttitle{Robust Pan-Cancer Mitotic Figure Detection with YOLOv12}

\author[1,2,3,4]{Rapha\"el Bourgade}
\author[1,5]{Guillaume Balezo}
\author[1]{Hana Feki}
\author[1,2,3]{Lily Monier}
\author[1,2,3]{Matthieu Blons}
\author[1,2,3]{Alice Blondel}
\author[4]{Delphine Loussouarn\textsuperscript{*}}
\author[2]{Anne Vincent-Salomon\textsuperscript{*}}
\author[1,2,3]{Thomas Walter}

\affil[1]{Centre for Computational Biology, MINES Paris--PSL University, Paris, France}
\affil[2]{Institut Curie, PSL University, Paris, France}
\affil[3]{INSERM, U1331 Computational Oncology, Paris, France}
\affil[4]{Department of Pathology, University Hospital of Nantes, Nantes, France}
\affil[5]{Sanofi, Paris, France}
\affil[ ]{\textsuperscript{*}These authors contributed equally to this work.}

\maketitle

\begin{abstract}
Mitotic figures represent a key histoprognostic feature in tumor pathology, providing crucial insights into tumor aggressiveness and proliferation. However, their identification remains challenging, subject to significant inter-observer variability, even among experienced pathologists. To address this issue, the MItosis DOmain Generalization (MIDOG) 2025 challenge marks the third edition of an international competition aiming to develop robust mitosis detection algorithms. In this paper, we present a mitotic figure detection approach based on the state-of-the-art YOLOv12 object detection architecture. Our method achieved an $F_1$-score of 0.801 on the preliminary test set (hot spots only) and ranked second on the final test leaderboard with an $F_1$-score of 0.7216 across complex and heterogeneous whole-slide regions, without relying on external data.
\end{abstract}


\begin{keywords}
MIDOG | Mitotic figure | YOLOv12 | Object detection
\end{keywords}

\begin{corrauthor}
raphael.bourgade@gmail.com
\end{corrauthor}

\section*{Introduction}
Detecting mitotic figures (MFs) in histopathology images remains a challenging task. Their quantification traditionally relies on the manual identification of ``hot spots'' by pathologists, followed by visual counting—an approach that is inherently subjective and may not reliably reflect the true proliferative activity of a tumor. With the rise of digital pathology and artificial intelligence, numerous efforts have been made to automate mitosis detection in order to enhance accuracy, reproducibility, and scalability. Among these, the \textit{MItosis DOmain Generalization} (MIDOG) challenges have emerged as a key benchmark for evaluating the generalizability of detection algorithms under realistic domain shifts. The 2021 edition~\cite{midog2021} addressed scanner-induced variability using breast cancer WSIs, while the 2022 edition~\cite{midog2022} extended the scope to include multiple tissue types and species, introducing further biological diversity. The 2025 MIDOG challenge~\cite{ammeling2025mitosis} builds on these foundations with the most comprehensive mitosis-annotated dataset to date, and introduces two tasks: (1) detecting mitotic figures in arbitrary tumor tissue, and (2) determining whether a mitotic figure is atypical or normal. These tasks represent a significant step toward developing robust mitosis detection systems that generalize across diverse and complex histological conditions. In this work, we present a high-performance detection pipeline based on the YOLOv12 object detection architecture.

\section*{Material and Methods}

\subsection*{Datasets}
We utilized all three datasets made available by the challenge organizers. The MIDOG++ dataset comprises 503 manually selected regions of interest (ROIs), each sampled at a resolution of 0.25~$\mu$m/px and measuring approximately $7000 \times 5000$ pixels. It encompasses seven tumor types from both human and canine origins: human breast carcinoma, canine lung carcinoma, canine lymphosarcoma, canine cutaneous mast cell tumor, human neuroendocrine tumor, canine soft tissue sarcoma, and human melanoma. This wide biological and technical diversity makes MIDOG++ a particularly valuable resource for training and evaluating mitosis detection models that are robust to inter-species variation, tissue-specific morphology, and scanner-induced artifacts.

In addition to MIDOG++, we incorporated two supplementary canine datasets released for the competition. The \textit{Canine Mammary Carcinoma} (CMC) dataset includes 21 fully annotated whole-slide images (WSIs) of canine breast carcinoma, while the \textit{Canine Cutaneous Mast Cell Tumor} (CCMCT) dataset contains 32 annotated WSIs of canine mast cell tumors. These datasets include 13{,}907 and 44{,}880 mitotic figure annotations, respectively. Due to the high mitotic activity typically observed in these tumor types, both datasets provide rich supervision for model training and were instrumental in improving detection performance.

\subsection*{Pre-processing}

We began by processing the two canine datasets through tissue segmentation, followed by tiling into approximately 1{,}800 regions of interest (ROIs) per dataset, each measuring around $6100 \times 5800$ pixels—matching the average ROI size found in the MIDOG++ dataset. To prepare the data for model input, we further subdivided each ROI into $640 \times 640$ pixel tiles using a 160-pixel overlap. This overlap was chosen specifically to reduce the risk of mitotic figures (MFs) being truncated or lying at tile borders, which could hinder detection accuracy.

All resulting tiles without any mitotic annotations were labeled as background. The final training dataset included 184{,}000 tiles sampled from the MIDOG++, MITOS CMC, and MITOS CCMCT datasets. To enhance the model’s discriminative capacity and ensure a balanced learning signal between mitotic and non-mitotic regions, we supplemented the training set with an additional 80{,}000 background tiles—randomly sampled from tissue areas devoid of mitotic figures.

\subsection*{Training}
We initially trained a one-stage YOLOv12-\emph{m} model \cite{tian2025yolov12} to simultaneously detect mitotic figures and challenging hard negative instances, using a 5-fold cross-validation scheme applied to the official training set of the MIDOG++ dataset. Each fold was trained independently, and the five resulting models were subsequently evaluated on the official test set provided by the challenge organizers. To produce final predictions, we averaged the detection metrics across the five folds. This procedure was designed to ensure strict consistency with the evaluation protocol of the original MIDOG++ benchmark \cite{aubreville2023comprehensive}, thereby enabling direct and reproducible comparison with previously published results.

\vspace{0.5em}

Following cross-validation, once the optimal set of hyperparameters was identified, we trained a final YOLOv12-\emph{m} model from scratch on the entire MIDOG++ dataset, combining both the original training and test sets to maximize the available annotated data. To further enhance the model’s generalization capabilities across diverse histological and acquisition domains, we extended training to include additional datasets—MITOS CMC and MITOS CCMCT—using the same preprocessing pipeline as previously described. These datasets introduce substantial variability in tissue morphology, staining characteristics, and scanning conditions, thereby exposing the model to a broader spectrum of mitotic and non-mitotic patterns. This extended training strategy was designed to improve robustness and domain transferability, particularly in challenging out-of-distribution scenarios.

\vspace{0.5em}

Training was performed for 50,000 iterations with a batch size of 64, using stochastic gradient descent (SGD) with an initial learning rate of $0.01$ and a weight decay of $5 \times 10^{-4}$. Momentum was set to 0.937, and we employed a warm-up strategy, with momentum and bias learning rate set to 0.8 and 0.1 respectively.
We applied data augmentations during training to increase robustness and diversity. These included horizontal and vertical flipping ($p = 0.5$), small random rotations between $-10^\circ$ and $+10^\circ$, and more contextual augmentation with mosaic augmentation ($p = 0.1$). In contrast, we explicitly disabled MixUp and CutMix, which could disrupt the spatial integrity of mitotic structures. No affine transformations such as translation, scale, shear, or perspective were applied, as these could distort the morphological features of mitotic figures and hinder the model’s ability to localize them accurately. Additionally, we used a custom batch sampler to mitigate domain imbalance, ensuring each batch included 50\% human and 50\% canine samples.
To simulate natural H\&E staining variability, we applied a multi-target Macenko stain normalization strategy \cite{ivanov2025multitarget}, extending classical Macenko by combining multiple reference stain profiles into a consensus transformation. We constructed a bank of 50 distinct normalization matrices, each derived from a different MIDOG++ ROI. During training, each tile had a 25\% chance of being normalized using one of these transforms, randomly selected at runtime. This stochastic augmentation strategy exposed the model to a wide range of staining conditions, enhancing its stain-invariant learning capacity.

\begin{figure}
\centering
\includegraphics[width=\linewidth]{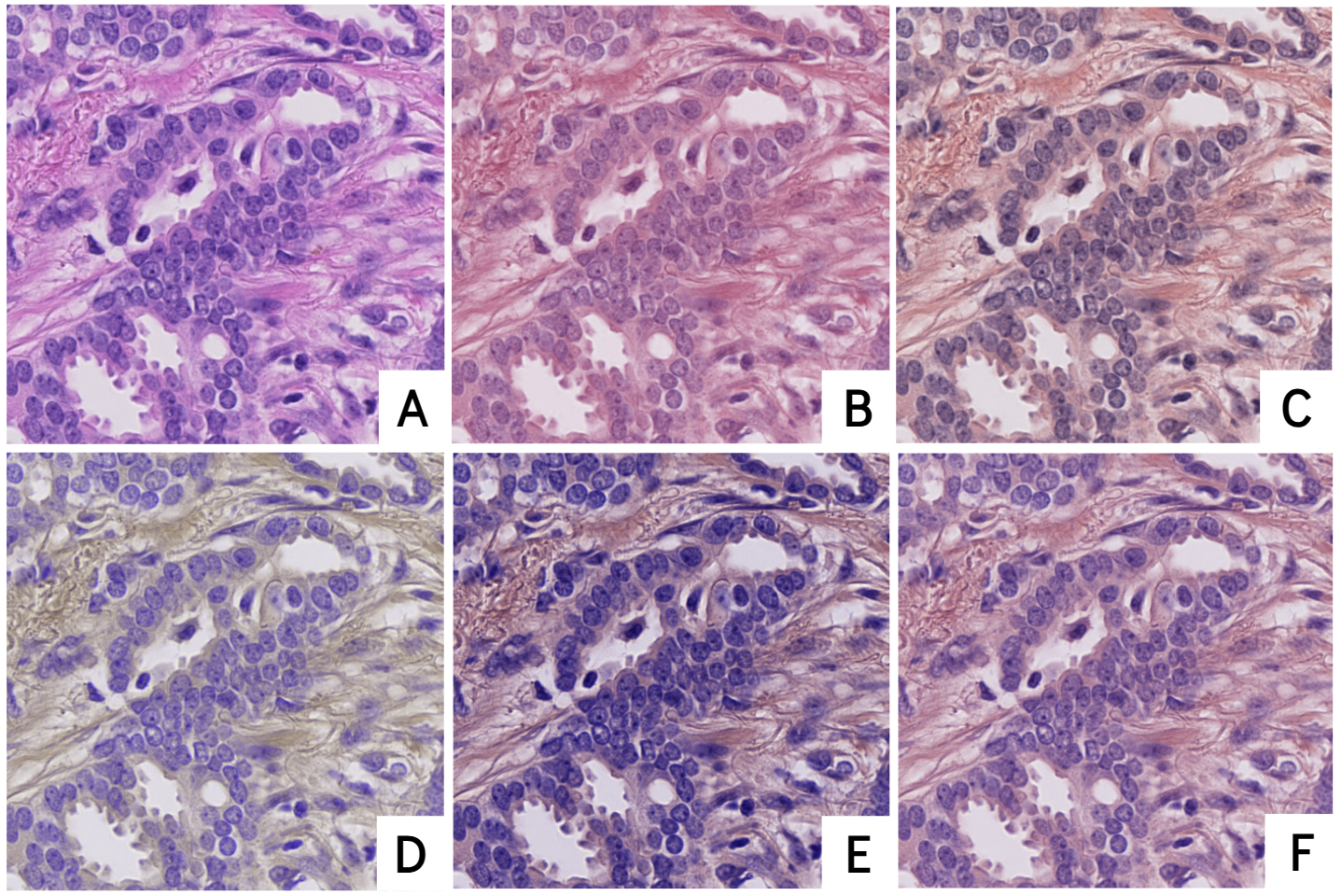}
\caption{Illustration of the augmentation workflow. (A) Example of a raw $640 \times 640$ H\&E tile extracted from an ROI. (B–F) Stain-normalized variants generated using a multi-target Macenko normalizer, mimicking natural staining variability across domains.}
\label{fig:computerNo}
\end{figure}

\subsection*{Post-processing}
At test time, each region of interest (ROI) was divided into overlapping patches of size 640$\times$640 pixels with a stride of 480 pixels, matching the training configuration. Each patch was processed independently by the YOLOv12 detector, using a Non-Maximum Suppression (NMS) threshold of 0.7 to filter overlapping predictions.

To project detections back into the original ROI coordinate space and account for patch overlaps, we implemented a dedicated post-processing pipeline. To improve robustness to image orientation and enhance recall, we performed Test-Time Augmentation (TTA): each tile was evaluated under a set of geometric transformations (horizontal/vertical flips), and the resulting predictions were merged.

To further consolidate redundant detections arising from patch overlaps and TTA, we applied Weighted Boxes Fusion (WBF) \cite{solovyev2021weighted}. This algorithm combines overlapping bounding boxes by computing a confidence-weighted average of their coordinates, effectively reducing false positives and improving spatial localization accuracy.

\section*{Results}

Our initial models, trained using 5-fold cross-validation on the official MIDOG++ training set, achieved a mean $F_1$-score of 0.817 and a mean Average Precision (mAP) of 0.867 when evaluated on the MIDOG++ test set (see Table~\ref{tab:combined-f1-ap}). These results demonstrate strong generalization across tumor types and confirm the robustness of our pipeline under intra-dataset variability.

\begin{table}[ht]
\centering
\scriptsize
\setlength{\tabcolsep}{6pt}
\renewcommand{\arraystretch}{1.3}
\begin{tabular}{|c|c|c|c|}
\hline
\textbf{Tumor Type} & \textbf{Metric} & \textbf{Baseline} & \textbf{Ours} \\
\hline
\multirow{2}{*}{\makecell{Breast carcinoma\\(human)}}           
& F1-score & $0.71 \pm 0.02$ & $\mathbf{0.798 \pm 0.01}$ \\
& AP       & $0.74 \pm 0.02$ & $\mathbf{0.841 \pm 0.02}$ \\
\hline
\multirow{2}{*}{\makecell{Lung carcinoma\\(canine)}}            
& F1-score & $0.68 \pm 0.02$ & $\mathbf{0.732 \pm 0.02}$ \\
& AP       & $0.65 \pm 0.06$ & $\mathbf{0.767 \pm 0.02}$ \\
\hline
\multirow{2}{*}{\makecell{Lymphosarcoma\\(canine)}}             
& F1-score & $0.73 \pm 0.01$ & $\mathbf{0.823 \pm 0.01}$ \\
& AP       & $0.69 \pm 0.04$ & $\mathbf{0.889 \pm 0.02}$ \\
\hline
\multirow{2}{*}{\makecell{Cutaneous Mast Cell\\(canine)}}       
& F1-score & $0.82 \pm 0.01$ & $\mathbf{0.890 \pm 0.01}$ \\
& AP       & $0.82 \pm 0.01$ & $\mathbf{0.938 \pm 0.02}$ \\
\hline
\multirow{2}{*}{\makecell{Neuroendocrine\\tumor (human)}}       
& F1-score & $0.59 \pm 0.01$ & $\mathbf{0.699 \pm 0.02}$ \\
& AP       & $0.62 \pm 0.03$ & $\mathbf{0.710 \pm 0.02}$ \\
\hline
\multirow{2}{*}{\makecell{Soft Tissue Sarcoma\\(canine)}}       
& F1-score & $0.69 \pm 0.01$ & $\mathbf{0.782 \pm 0.01}$ \\
& AP       & $0.69 \pm 0.03$ & $\mathbf{0.824 \pm 0.03}$ \\
\hline
\multirow{2}{*}{\makecell{Melanoma\\(human)}}                  
& F1-score & $0.81 \pm 0.01$ & $\mathbf{0.826 \pm 0.01}$ \\
& AP       & $0.82 \pm 0.03$ & $\mathbf{0.869 \pm 0.02}$ \\
\hline
\multirow{2}{*}{\textbf{Mean across types}}                   
& F1-score & Not provided  & $\mathbf{0.817 \pm 0.02}$ \\
& AP       &  Not provided & $\mathbf{0.867 \pm 0.03}$ \\
\hline
\end{tabular}
\caption{Comparison of F1-score and Average Precision (AP) $\pm$ standard deviation across 5-fold cross-validation for each tumor type in the MIDOG++ challenge. The reported mean corresponds to the average computed across all individual test images. Baseline results are from Aubreville et al. (2023)~\cite{aubreville2023comprehensive}.}
\label{tab:combined-f1-ap}
\end{table}

Subsequently, a final YOLOv12-\emph{m} model was trained from scratch on the entire training and test sets of MIDOG++, augmented with the MITOS CMC and MITOS CCMCT datasets to maximize the diversity and volume of annotated data. This model achieved an $F_1$-score of 0.801 on the MIDOG 2025 preliminary test set, which consists exclusively of annotated hot spot regions. The associated precision and recall were 0.808 and 0.794, respectively, indicating a balanced operating point and consistent detection performance.

Finally, we evaluated this model on the MIDOG 2025 final test set, which included more complex tissue contexts beyond hot spots—such as inflamed, necrotic, and non-tumor areas, as well as randomly sampled regions. On this more challenging benchmark, our approach achieved an $F_1$-score of 0.722, with a precision of 0.773 and a recall of 0.677. These results positioned our method 2\textsuperscript{nd} overall on the competition leaderboard, underscoring its robustness and adaptability to real-world variability.

\section*{Discussion}

Our results demonstrate that a single-stage detector such as YOLOv12, when coupled with domain-aware data augmentation through stain normalization and an effective tile-to-ROI level fusion strategy, can deliver robust performance—even on ROIs originating from previously unseen domains. Notably, our pipeline avoids both external data and model ensembling, which significantly reduces system complexity, accelerates inference, and enhances deployability in real-world clinical settings.

Despite these strengths, our approach is not without limitations. It relies on fixed heuristic thresholds for confidence scoring and post-processing (e.g., NMS and WBF), which may be suboptimal in the face of domain or scanner shifts. These thresholds directly influence the precision–recall balance and could potentially degrade performance under strong out-of-distribution conditions.

To explore whether further gains could be achieved, we experimented with a two-stage detection-classification pipeline. In this setup, YOLOv12 first generated candidate detections with confidence scores ranging from 0 to 1, which were then refined by a ViT-H+ classifier. The classifier operated on 128~$\times$~128 context windows centered around each detection and was trained using the DINOv3 self-supervised strategy \cite{simeoni2025dinov3} on the combined MIDOG++, MITOS CMC, and MITOS CCMCT datasets. Its goal was to distinguish true mitotic figures from visually similar hard negatives. However, in our experiments, this two-stage pipeline failed to outperform the single-stage YOLOv12 detector. One likely explanation is that the initial proposals generated by YOLOv12 were already of high quality—leaving limited room for the second-stage classifier to improve upon them, especially given that both stages were trained on overlapping annotations.

In summary, this work introduces a high-performing and computationally efficient single-stage mitosis detection pipeline based on YOLOv12, enhanced through domain-aware stain normalization and robust post-processing techniques. Our approach achieved competitive results on the MIDOG 2025 benchmark without the need for external data or ensemble models, thus offering a reproducible and clinically feasible solution.

Looking ahead, future research should investigate the integration of large-scale, curated external datasets to introduce additional diversity in staining, tissue morphology, and scanner characteristics—thereby improving domain generalization. Additionally, retraining the second-stage classifier on more heterogeneous and difficult examples—particularly false positives identified by the detector—could enhance its discriminative capabilities and mitigate the limitations of fixed confidence thresholds in YOLO-based inference.

\begin{acknowledgements}
The authors gratefully acknowledge the MIDOG organizers and the Grand-Challenge platform for hosting the challenge and providing access to such valuable datasets.

This work was supported by the French government under the management of Agence Nationale de la Recherche as part of the “Investissements d’avenir” program, reference ANR-19-P3IA-0001 (PRAIRIE 3IA Institute). Furthermore, this work was supported by a government grant managed by the Agence Nationale de la Recherche under the France 2030 program, with the reference numbers ANR-24-EXCI-0001, ANR-24-EXCI-0002, ANR-24-EXCI-0003, ANR-24-EXCI-0004, ANR-24-EXCI-0005.

\end{acknowledgements}

\section*{Bibliography}
\bibliography{literature}

\end{document}